\pdfoutput=1
\documentclass[journal]{IEEEtran}
\usepackage{amsmath}
\usepackage{multirow}
\usepackage[scaled=1.0]{helvet}
\usepackage{times}
\usepackage{graphicx}
\usepackage{subfigure}
\usepackage{parskip}
\usepackage{color}
\usepackage[labelfont=bf,textfont=it]{caption}

\begin{document}
\title{Exposure Interpolation by Combining Model-driven and Data-driven Methods}
%
%
%
\author{Chaobing Zheng, Zhengguo Li, and Shiqian Wu$^*$
\thanks{*  Corresponding author}
\thanks{Chaobing Zheng and Shiqian Wu are with the Institute of Robotics and Intelligent Systems, school of Information Science and Engineering, Wuhan University of Science and Technology, Wuhan 430081, China(e-mails: zhengchaobing@wust.edu.cn, shiqian.wu@wust.edu.cn).}
\thanks{Zhengguo Li is with the SRO department, Institute for Infocomm Research, Singapore, 138632, (email: ezgli@i2r.a-star.edu.sg).}}
%
%

\markboth{}
{Shell \MakeLowercase{\textit{et al.}}: Bare Demo of IEEEtran.cls
for Journals}
%



\maketitle

\begin{bfseries}
\begin{abstract}
Brightness order reversal could happen among over-exposed regions of a bright image and under-exposed regions of a dark image if two large-exposure-ratio images are fused directly by using existing multi-scale exposure fusion (MEF) algorithms. This problem can be addressed efficiently by interpolating a virtual image with a medium exposure time. In this paper, a new exposure interpolation algorithm is introduced by combining model driven and data driven exposure interpolation methods. The key idea is to obtain an initial medium-exposure image by using intensity mapping functions (IMFs), while the modeling error is compensated by the data-driven method. Experimental results indicate that the data-driven method is benefited from the model-driven method for fast convergence speed and demand of large training samples. The final interpolated medium-exposure image is significantly improved by employing the hybrid methods in terms of PSNR and SSIM metrics.
\end{abstract}

\end{bfseries}

%
\IEEEpeerreviewmaketitle

\section{Introduction}

Due to limitations of existing digital device sensor,  combining differently exposed images to expand the dynamic range is a simple method to obtain an image with more information \cite{1debevec1997}. Existing multi-scale exposure fusion (MEF) algorithms \cite{1mertens2007,ZGLi2017,Ancuti2017,kou2017}  assume that there is neither camera movement nor moving objects in all the differently exposed images. The assumption is not true if all the differently exposed images are captured by using the method in \cite{1debevec1997}. The fused image is blurred if there are camera movements and there are ghosting artifacts if there are moving objects. It is not difficult to align the differently exposed images \cite{1wu2014} but it is very challenging to synchronize all the moving objects in the differently exposed images \cite{1zheng2013}. As such, ghosting artifacts are believed to be the Achilles' Heel for existing high dynamic range (HDR) imaging solutions.

\begin{figure}[htb]
	\centering
	\includegraphics[width=0.35\textwidth]{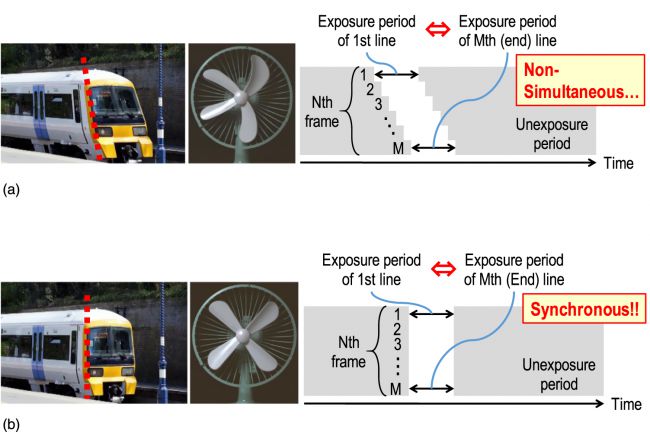}
	\caption{Skewing artifacts as recorded by a standard rolling shutter (a), and as eliminated by a global shutter (b), image courtesy of \cite{1cannon2019}.}
	\label{figure1}
\end{figure}

New HDR image capturing devices are introduced to address the above problems. One example is a beam splitting based HDR video capturing system with few sensors \cite{1tocci2011}. The number of sensors can be reduced to two in order to save the cost. Another one is a row-wise CMOS HDR video capturing system \cite{1Gu2010}. An image is split into two fields with differently exposed times to simplify the CMOS sensor. The rolling-shutter suffers from skewing as shown in Fig. \ref{figure1}. It is seen from Fig. \ref{figure1} that if there is any moving object, then the data which is recorded by the lower half of the sensor will be in a slightly different position. Recently, the Canon released an innovative global shutter with a specific sensor that reads the sensor twice in an HDR mode \cite{1cannon2019}. All these devices can also be applied to capture HDR videos which will be more and more popular in the coming 5G era.

The ratio between the exposure times could be very large so as to capture information as much as possible from an HDR scene. Since shadow regions in the bright image could be darker than high-light regions in the dark image, the MEF methods \cite{1mertens2007,ZGLi2017,Ancuti2017,kou2017} could suffer from brightness order reversal among the shadow regions in the bright image and  high-light regions in the dark image \cite{1yang2018}. The fused image will look unnatural. Halo artifacts could also be produced in the fused image. Exposure interpolation is an effective way to address the problem  as shown in \cite{1yang2018}. The intensity mapping functions (IMFs) between a pair of differently exposed images are calculated, by which a medium-exposure image is generated \cite{1yang2018}. However, the limited representation capability of the IMFs results in a low quality medium-exposure image which will affect the quality of finally fused image \cite{1zheng2020}.

In view of limitation of the IMF based algorithm in \cite{1yang2018, 1zheng2020} and much stronger representation capability of data-driven methods, fusing model-driven and data-driven methods might be an efficient way for the exposure interpolation \cite{1zheng2020}. This is elaborated by borrowing wisdom from the field of nonlinear control system. Modelled dynamics and unmodeled dynamics are two well known concepts in field of nonlinear control systems \cite{1khalil2002}. Inspired by this idea, two new concepts, modelled information and unmodeled information are introduced to design a hybrid framework on fusing model-driven and data-driven methods here. Assuming the exposure interpolation of two large-exposure-ratio images $x_1$ and $x_2$ \cite{1yang2018, 1zheng2020}, and the ground truth of the medium exposure image be denoted as $y$, the relationship between $x_1$, $x_2$ and $y$ is usually represented by a nonlinear equation $y=f(x_1,x_2)$. Using the method in \cite{1yang2018, 1zheng2020}, an intermediate medium exposure image $y_0$ can be obtained as $y_0=f_0(x_1, x_2)$. Here, $f_0(x_1, x_2)$ is the modelled information $y$ by the method in \cite{1yang2018, 1zheng2020} and $(y-y_0)$ is unmodeled information by the method in \cite{1yang2018, 1zheng2020} with respect to $y$. Clearly, the quality of the virtually medium exposure image can be improved if the unmodeled information can be further represented. Fortunately,  the unmodeled information can be represented by a deep convolutional neural network (CNN) such as DenoiseNet \cite{CycleISP}. This implies that a deep learning method can be adopted to improve the conventional method. It is thus desired to develop a new exposure interpolation algorithm by fusing the model-driven and data-driven approaches.

In this paper, a new hybrid exposure interpolation framework is introduced to fuse a model-driven exposure interpolation method with a data driven based exposure interpolation method. In other words, this paper intends to explore the feasibility of {\it compensating} a model-driven image processing method with a data-driven image processing method rather than a sophisticated neural network for deep learning. Specifically, an intermediate image $y_0$ is firstly produced by using new IMFs which outperforms the IMFs in \cite{1yang2018, 1zheng2020}. Unmodeled (or residual) information $(y-y_0)$ is thus less than that in \cite{1zheng2020}. Unlike the data-driven method in the single image brightening in \cite{A} which is supposed to hallucinate information in under-exposed regions, noise reduction is the main task of the data-driven method in the proposed exposure interpolation. The DenoiseNet \cite{CycleISP} is therefore selected to approximate the unmodeled information $(y-y_0)$ via a supervised learning approach, which differs fundamentally from existing data-driven approaches. The DenoiseNet have several recursive residual groups (RRG) which contain multiple dual attention blocks (DAB). Each DAB contains spatial attention and channel attention modules, which can suppress the less useful features and only allow the propagation of more informative ones. Compared with an existing data-driven method which uses a CNN to approximate $y$ directly, the proposed framework reduces the amount of training data and improves the convergence speed. This is not surprised because the residual image $(y-y_0)$ is much sparser than the image $y$. Meanwhile, the quality of the intermediate image $y_0$ is significantly improved due to compensating unmodeled error by the deep learning method. The peak signal to noise ratio (PSNR) and the structural similarity index (SSIM) of the resultant fused images are on average have improved much, respectively. Clearly, the model-driven method and the data-driven method {\it compensate} each other in the proposed hybrid exposure interpolation framework. To validate the necessity of exposure interpolation, the interpolated image and two large-exposure-ratio images are fused together via the MEF algorithm in \cite{1mertens2007}. Experimental results indicate that the resultant MEF algorithm outperforms the five state-of-the-art MEF algorithms in \cite{1yang2018,1mertens2007,Ancuti2017,ZGLi2017,kou2017} when the inputs are the two large-exposure-ratio images. In addition, the possible relative brightness change is indeed overcome by the proposed exposure interpolation algorithm. The halo artifacts are also significantly reduced. In summary, the contributions are highlighted as follows:

 1) A database which consists of 370  multi-exposed image sequences has been built up. To avoid other influences, only exposure time is changed while other configurations of the cameras are fixed. Camera shaking, object movement are strictly controlled to ensure that only illumination is changed.

 2) A hybrid $L_1/L_2$ loss function is proposed. The new function outperforms the popular $L_1$ and $L_2$ loss functions.

 3) A hybrid framework is introduced in this paper. The model driven and data-driven methods {\it compensate} each other in the proposed framework.  A new exposure interpolation algorithm is designed by using the hybrid framework. The algorithm can be used to improve the performance of existing MEF algorithms when inputs are two-large-exposure-ratio images.

The rest of this paper is organized as follow: A hybrid exposure interpolation framework is introduced in Section \ref{paradigm}.  Experimental result are provided in Section \ref{experimentalresult} to verify the proposed framework. Finally, conclusions are drawn in Section \ref{conclusion}.

\begin{figure*}[htb]
	\centering
	\subfigure{
		\begin{minipage}[b]{0.225\linewidth}
			\includegraphics[width=1\linewidth]{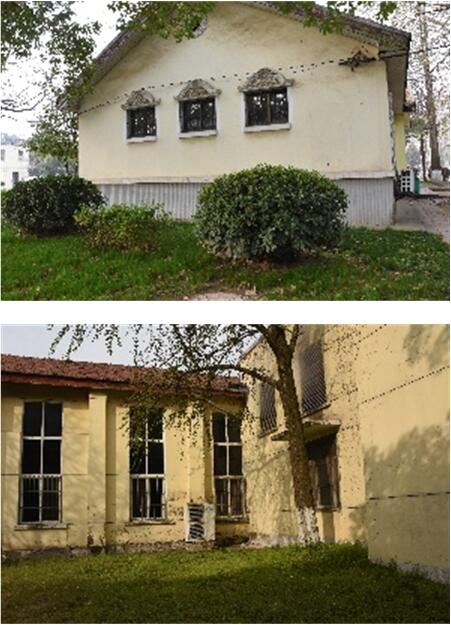}	
			\centerline{(a)}	
	\end{minipage}}
	\subfigure{
		\begin{minipage}[b]{0.225\linewidth}
			\includegraphics[width=1\linewidth]{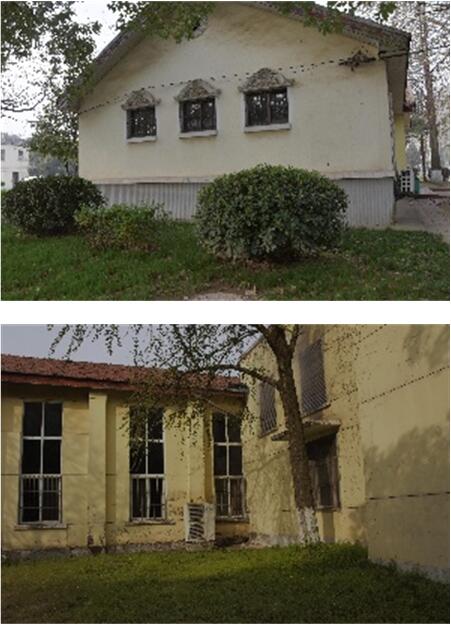}
			\centerline{(b)}
	\end{minipage}}
	\subfigure{
		\begin{minipage}[b]{0.331\linewidth}
				\includegraphics[width=1\linewidth]{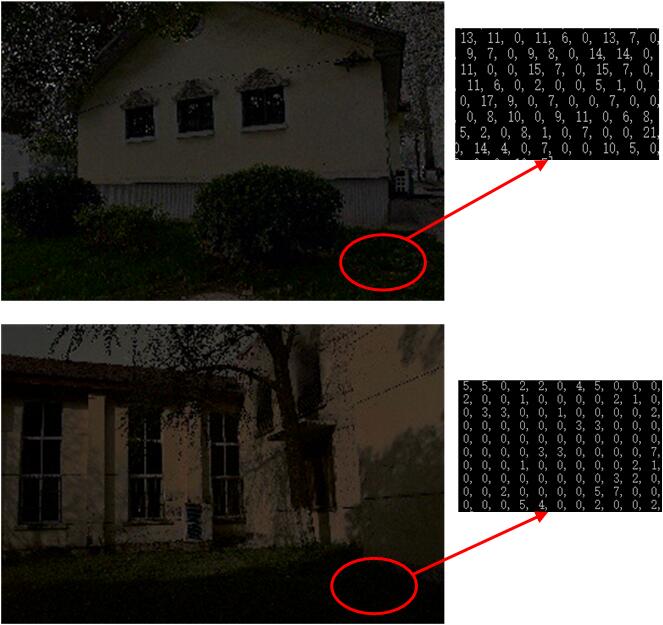}
			\centerline{(c)}
	\end{minipage}}
	\caption{(a) the ground truth images $y$; (b) the intermediate images $y_0$; (c) unmodeled information $(y-y_0)$. The unmodeled information is usually small, many pixel values are 0's.}
\label{fig23}
\end{figure*}

\section{Exposure Interpolation via A Hybrid Framework}
\label{paradigm}

In this section, a hybrid framework is introduced for exposure interpolation. The framework is composed of a model-driven exposure interpolation method and a data-driven based exposure interpolation method. They {\it compensate} each other.

\subsection{The Proposed  Hybrid Framework}

Let $x_1$ and $x_2$ be two large-exposure-ratio images of the same scene. The exposure times are $\Delta t_1$ and $\Delta t_2$, respectively. Without loss of generality, $\Delta {t_1}\gg\Delta {t_2}$. Let $y$ be the ground-truth image of the medium-exposure image. The exposure time of $y$ is assumed between $\Delta {t_1}$ and $\Delta {t_2}$ which is defined as:
\begin{equation}
\label{eq3}
\Delta {t_3} = \sqrt {\Delta t_1\Delta t_2}.
\end{equation}

A  data-driven based exposure interpolation method intends to use a  deep CNN to represent $y$ by
\begin{equation}
y= f(x_1,x_2).
\end{equation}

Convergence of the method is an important issue. Many different methods were provided to address this issue and good examples are given in \cite{1he2016, Huang2016,Kim}. A new hybrid framework will be proposed in this section to address the issue.

Inspired by the concepts of modelled dynamics and unmodelled dynamics in the field of nonlinear control systems \cite{1khalil2002},  $f(x_1,x_2)$ can be decomposed as
\begin{equation}
f(x_1,x_2)=f_0(x_1,x_2)+\tilde{f}(x_1,x_2),
\end{equation}
where $f_0(x_1,x_2)(\doteq y_0)$ is an initial representation of $y$ which is obtained using a conventional exposure interpolation method such as  \cite{1yang2018}. $y_0$ and $\tilde{f}(x_1,x_2)$ can be regarded as modelled information and  unmodelled (or remaining) information of $y$ with respect to the conventional exposure interpolation method, respectively.

 Let $(y-y_0)$ be denoted as $\tilde{y}$ which can be regarded as unmodeled information of $y$. Let $\|y\|_0$ be the number of non-zeros in the image $y$. Normally, $\|\tilde{y}\|_0$ is smaller than $\|y\|_0$. One example is given in Fig. \ref{fig23}. In other words, $\tilde{y}$ is sparser than $y$. In addition, $\|\tilde{y}\|_1$ is smaller than $\|y\|_1$.

 Instead of training a CNN as in the existing deep learning to approximate  $y$, a new CNN is trained to approximate $\tilde{y}$.  It would be easier to train the latter CNN using a residual network \cite{1he2016}. It can be expected that the convergence of the new CNN would be increased while the number of training samples would be reduced.

\begin{figure}[htb]
	\centering
	\includegraphics[width=0.48\textwidth]{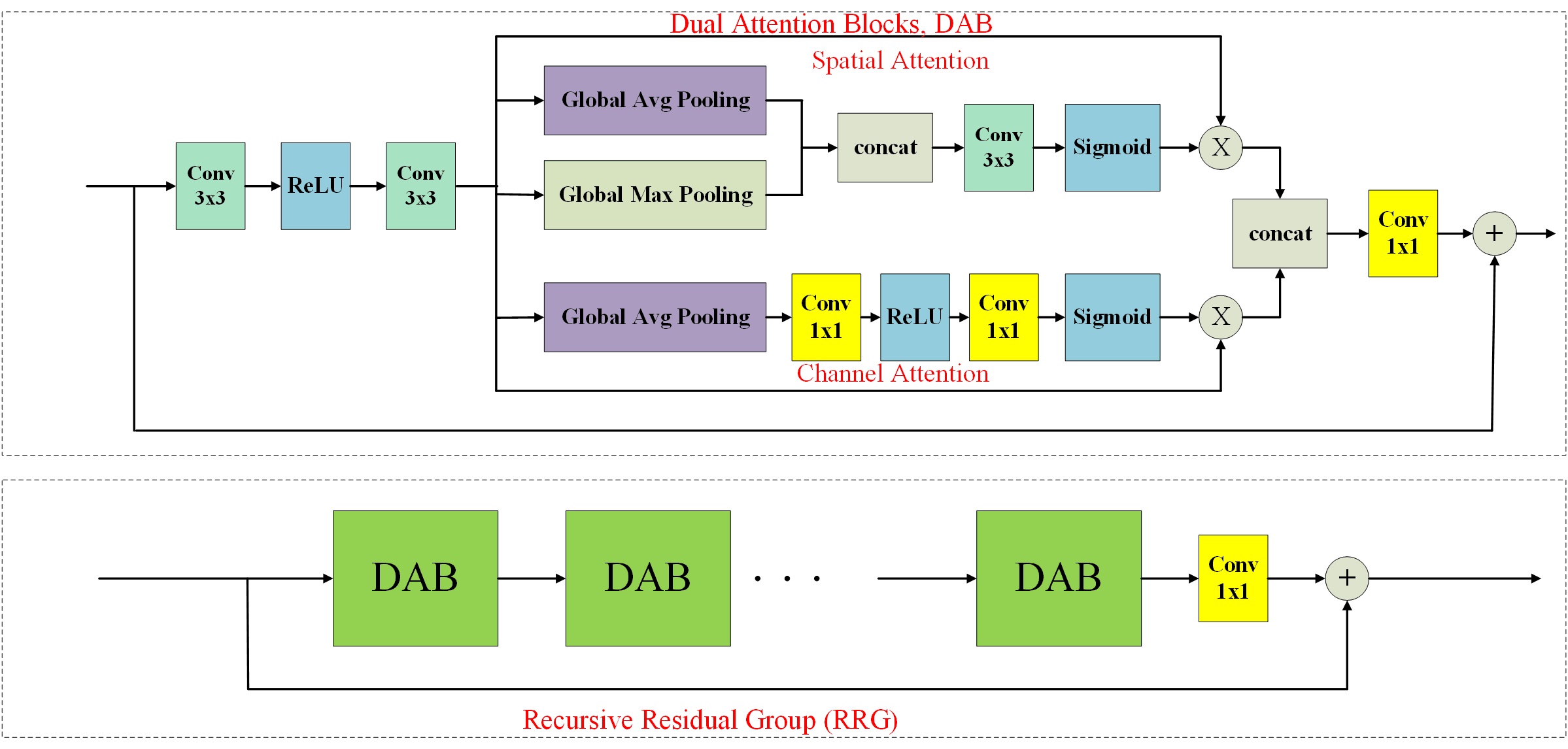}
	\caption{Recursive residual group (RRG) contains multiple dual attention blocks (DAB) \cite{CycleISP}. Each DAB contains spatial attention and channel attention modules.}
	\label{Fig2}
\end{figure}

\begin{figure*}[htb]
	\centering
	\includegraphics[width=0.95\textwidth]{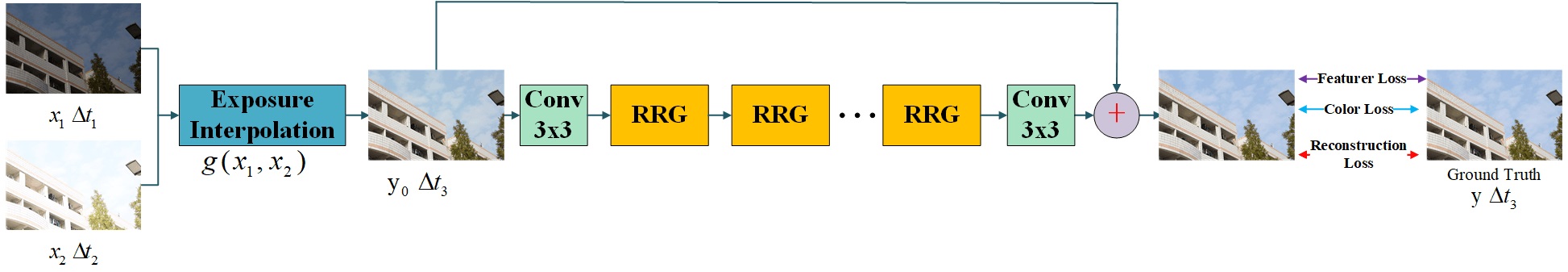}
	\caption{A hybrid framework for exposure interpolation. An intermediate image $y_0$ is first produced by a proposed model-driven method, DenoiseNet \cite{CycleISP} is then trained to learn $(y-y_0)$ from two images $\left\{ y, y_0 \right\}$ with the proposed loss function.}
	\label{Fig.123}
\end{figure*}

\begin{figure*}[!htb]
	\centering
	\includegraphics[width=0.95\textwidth]{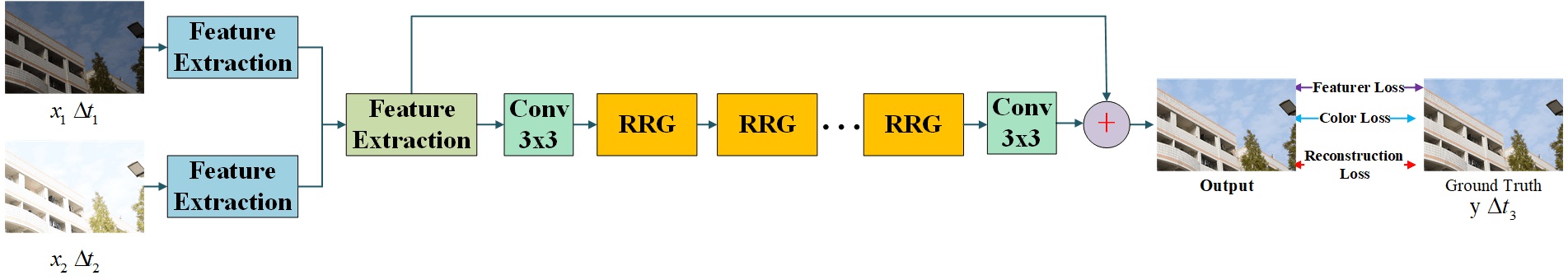}
	\caption{A data-driven method for exposure interpolation. Firstly, $x_1$ and $x_2$ are processed  by two different convolutional neural network to generate the corresponding features. Then, the concatenated features are further processed by a convolutional neural network. Loss function is similar to the proposed approach. }
	\label{Fig6}
\end{figure*}

Fig. \ref{Fig.123} summarizes the proposed hybrid framework exposure interpolation via fusing a model-driven and a data-driven methods while Fig. \ref{Fig6} shows an existing data-driven approach. Clearly, the proposed hybrid framework is fundamentally different from the existing data-driven approach in the sense that the proposed framework learns the $\tilde{y}=(y-y_0)$ so as to approach the ground truth image. According to the proposed framework, an intermediate image will be firstly generated using the  method in \cite{1yang2018}. A data-driven based method will then be designed to refine the intermediate image. The details are provided in the following two subsections.

\subsection{Generation of Intermediate Image $y_0$} \label{section:IMF}
The intermediate image is generated by finding the relationships between the interpolated image and the two large-exposure-ratio images. Assume the camera response functions (CRF) be $F_c(\cdot)$. Here, $c\in\{R, G, B\}$ is a color channel. Let the intensity mapping functions (IMF) from $x_{1,c}$ to $y_{0,c}$ and from $x_{2,c}$ to $y_{0,c}$ be denoted as ${\Lambda_{1,3,c}}(\cdot)$ and $\Lambda_{2,3,c}$, respectively \cite{Grossberg2003}.  The functions ${\Lambda_{1,3,c}}(\cdot)$ and ${\Lambda_{2,3,c}}(\cdot)$ can be expressed as:
\begin{align}
\label{eq1}
\Lambda_{i,3,c}(z) & = F_c(\frac{\Delta t_3}{\Delta t_i}F^{-1}_c(z))\; ;\; i\in \{1,2\}.
\end{align}

The $F_c^{-1}(z)$ maps an integer $z$ in $[0, 255]$ to the corresponding irradiance. $\frac{\Delta t_3}{\Delta t_i}F^{-1}_c(z)(=\tilde{z})$ is sometimes between two adjacent  mapped irradiance, and the corresponding pixel value cannot be directly obtained. Thus, it is necessary to estimate the function curve according to the known data points in advance.
Three curves with parameters $k_i(1\leq i\leq 6)$  are adopted in in \cite{1yang2018,1zheng2020} to fit the $F_c(\cdot)$ scatter plot as
\begin{align}
\label{fit}
z=\frac{{{k}_{1}}}{1+{{e}^{{{k}_{2}}+{{k}_{3}}\times \tilde{z}}}}+\frac{{{k}_{4}}}{1+{{e}^{{{k}_{5}}+{{k}_{6}}\times \tilde{z}}}}.
\end{align}

Although the fitting (\ref{fit}) is relatively smooth, it cannot guarantee that all data points calculated by (\ref{fit}) are on the curve which will affect the interpolated image. A more accurate linear interpolation method is adopted in this paper. The new interpolation method can ensure that all points to be interpolated are on the curve, the estimated $F_c(\cdot)$ curve is thus more accurate. Subsequently, the interpolated image is closer to the ground truth as shown in experimental results in the Section III.B.

Same as \cite{1yang2018}, two virtual images $\Lambda_{13}(x_1)$ and $\Lambda_{23}(x_2)$ are generated. As the IMF is incredible in when mapping a pixel in an under-exposed region of the dark image to a bright image, and it is also incredible  when mapping a pixel in an over-exposed region of bright image to a dark image \cite{1li2014}.  Therefore,
the intermediate image $y_0$ is generated by fusing them via the following formula:
\begin{equation}
\label{newfusion}
y_{0,c}(p) = \frac{\sum_{i=1}^2W_i(x_{i,c}(p))\Lambda_{i,3,c}(x_{i,c}(p))}{\sum_{i=1}^2W_i(x_{i,c}(p))},
\end{equation}
where the weighting functions $W_1(z)$ and $W_2(z)$ are defined as:
\begin{eqnarray}
&&\hspace{-7mm}W_1(z) = \left\{ {\begin{array}{*{20}{l}}
	{0;}&{if~}{{\rm{0}} \le z < {\xi _L}}\\
	{1 - 3{h_1}^2(z) + 2{h_1}^3(z);}&{if~}{{\xi _L} \le z < 55}\\
	{1;}&{otherwise}{}.
	\end{array}} \right.,\end{eqnarray}\begin{eqnarray}
&&\hspace{-7mm}W_2(z) = \left\{ {\begin{array}{*{20}{l}}
	{1;}&{if~}{{\rm{0}} \le z < 200}\\
	{1 - 3{h_2}^2(z) + 2{h_2}^3(z);}&{if~}{200 \le z < {\xi _U}}\\
	{0;}&{otherwise}{}.
	\end{array}} \right.,
\end{eqnarray}
and $h_1(z)$ and $h_2(z)$ are defined as
\begin{eqnarray}
&&\hspace{-5mm}
h_1(z)=\frac{55-z}{55-\xi_L},\\
&&\hspace{-7mm}h_2(z)=\frac{z-200}{\xi_U-200}.
\end{eqnarray}


Clearly, the generation of the intermediate image needs a low computational cost. Actually, the simplicity of the conventional methods is a very important criteria when the model-driven methods are fused with data-driven methods to address image processing problems.

$\tilde{y}(=y-y_0)$ is unmodeled information by the new fusion method (\ref{newfusion}).  In the next subsection, a data-driven method will be designed to represent the residual image $\tilde{y}$.

\subsection{Refinement of Intermediate Image $y_0$}
Unlike single image brightening in \cite{A} which restores details in the under-exposed regions via hallucination, noise reduction is the main issue for the exposure interpolation. The DenoiseNet in \cite{CycleISP} is thus selected to refine the intermediate image $y_0$. As mentioned in the introduction, the unmodeled information $\tilde{y}$ is sparser than the original information $y$, and most values are likely to be zero or small as shown in Fig. \ref{fig23}. It can be expected that it is easier to use a neural network to approximate $\tilde{y}$ than $y$. In this subsection, the DenoiseNet \cite{CycleISP} will be adopted to approximate $\tilde{y}$ as shown in Fig. \ref{Fig2} and \ref{Fig.123}. The DenoiseNet has two attractive characteristics: (1) The structure of DenoiseNet is a residual network. It is important to compress the mapping range during the training of the network \cite{residual}. It is much easier for the residual structure to learn the mapping. (2) Recursive Residual Group (RRG) is widely used in DenoiseNet as shown in Figs. \ref{Fig2} and \ref{Fig.123}. The RRG contains $n$ dual attention blocks (DAB). The goal of each DAB is to suppress the less useful features and only allow the propagation of more informative ones. Because two attention mechanisms channel attention (CA) and spatial attention (SA) are adopted to achieve this performance by the DAB.

Loss functions play an important role in the data-driven approach. The unmodeled information $\tilde{y}$ is learned from two images $\left\{y, y_0 \right\}$ by minimizing the following loss function:
\begin{equation}
\label{eq14}
L_d= L_r + w_cL_c+w_fL_f,
\end{equation}
where $w_c$ and $w_f$ are two constants, and their values are selected as 0.01 and 0.01, respectively if not specified in this paper. $L_r$ is the reconstruction loss function,  $L_c$ represents the color loss function, $L_f$ is feature-wise loss function.

Two choices for the reconstruction loss function $L_r$  are provided as
\begin{align}
\label{lossfunction1}
\left\{\begin{array}{l}
L_r = {\|\tilde{y}-\tilde{f}(y_0)\|}_1= {\|y-y_0-\tilde{f}(y_0)\|}_1\\
L_r = {\|\tilde{y}-\tilde{f}(y_0)\|}_2= {\|y-y_0-\tilde{f}(y_0)\|}_2
\end{array}
\right.,
\end{align}
and they are different from the following loss functions
\begin{equation}
\label{lossfunction2}
\left\{\begin{array}{l}
L_r = \|y-f(x_1,x_2)\|_1\\
L_r = \|y-f(x_1,x_2)\|_2
\end{array}
\right.,
\end{equation}
which are widely used in the existing data-driven based methods.

The new reconstruction loss function $L_r$ is a hybrid $L_1/L_2$ norm which is on top of the functions in Eq (\ref{lossfunction1}), and it is given as
\begin{equation}
\label{lrlr}
L_r = \sum_p\psi(y(p)-y_0(p)-\tilde{f}(y_0(p))),
\end{equation}
where the function $\psi(z)$ is defined as \cite{1wangwei2019}
\begin{equation}
\psi(z)=\left\{\begin{array}{ll}
|z|;&\mbox{if~}|z|>c\\
\frac{z^2+c^2}{2c}; &\mbox{otherwise}\\
\end{array}
\right.,
\end{equation}
and $c$ is a positive constant and its value is selected as 5/255 in this paper.

It is easily shown that the function  $\psi(z)$ is differentiable. Let  $\psi'(z)$ be the derivative
of the function $\psi(z)$, and it is clearly a continuous function given as:
\begin{equation}
\psi'(z)=\left\{\begin{array}{ll}
1; &\mbox{if~}z\geq c\\
-1; &\mbox{if~}z\leq -c\\
\frac{z}{c}; &\mbox{otherwise}\\
\end{array}
\right..
\end{equation}

It is noted that it may exist color distortion by using the restoration loss only  because $L_r$ metric measures the color difference numerically, and not produce correct details and vivid color,  as shown in Fig. \ref{Fig8a}, \ref{fig56}. Hence, one more color loss is introduced follows:
\begin{equation}
\label{a1}
L_c=\sum_p\angle (y(p), y_0(p)+\tilde{f}(y_0(p))),
\end{equation}
where $\angle(y(p),y_0(p)+\tilde{f}(y_0(p)))$ is the angle between two 3D $(R, G, B)$ vectors $y(p)$ and $ (y_0(p)+\tilde{f}(y_0(p)))$. Eq. (\ref{a1}) sums the angles between the color vectors for every pixel pair in the enhanced images $(y_0+\tilde{f}(y_0))$ and the ground truth images $y$. Such loss function ensures that the color vectors have the same direction and reduces the possible color distortion \cite{Endo,DeepUPE}.

Both the $L_r$ and the $L_c$ are the pixel-wise loss functions which accurately capture the low frequencies but fail to encourage high frequency crispness. The resultant virtual image is high fidelity but not realistic. The statements are too subjective. The image is  usually overly-smooth and thus has poor perceptual quality. Thus, feature-wise loss functions is applied to enhance the pixel-wise loss functions.  Instead of using commonly adopted feature-wise loss function that adopts a VGG network trained for image classification, a fine-tuned VGG network for material recognition in \cite{1bell2015} is adopted to define the feature-wise loss. The VGG in \cite{1bell2015} focuses on textures rather than object and the texture is critical for the refinement of the virtual image. The feature-wise loss $L_f$ is defined as
\begin{align}
L_f=\frac{1}{W_{i,j}}\frac{1}{H_{i,j}}\sum_{l=1}^{W_{i,j}}\sum_{m=1}^{H_{i,j}}(\phi_{i,j}(y)_{l,m}-\phi_{i,j}(y_0+\tilde{f}(y_0))_{l,m})^2,
\end{align}
where $W_{i,j}$ and $H_{i,j}$ denote  the dimensions of the respective feature maps within the VGG network. $\phi_{i,j}(\cdot)$ is the feature map obtained by the $j$-th convolution (before activation) before the $i$-th maxpooling layer within the VGG network.

\section{Experimental Results}
\label{experimentalresult}
Extensive experimental results are provided to validate the proposed hybrid framework with emphasis on illustrating how the model-driven method and the data-driven method {\it compensate} each other. Readers are invited to view to electronic version of full-size figures and zoom in these figures so as to better appreciate differences among images.

\subsection{Datesets}
\begin{figure*}[htb]
	\centering
	\includegraphics[width=1\textwidth]{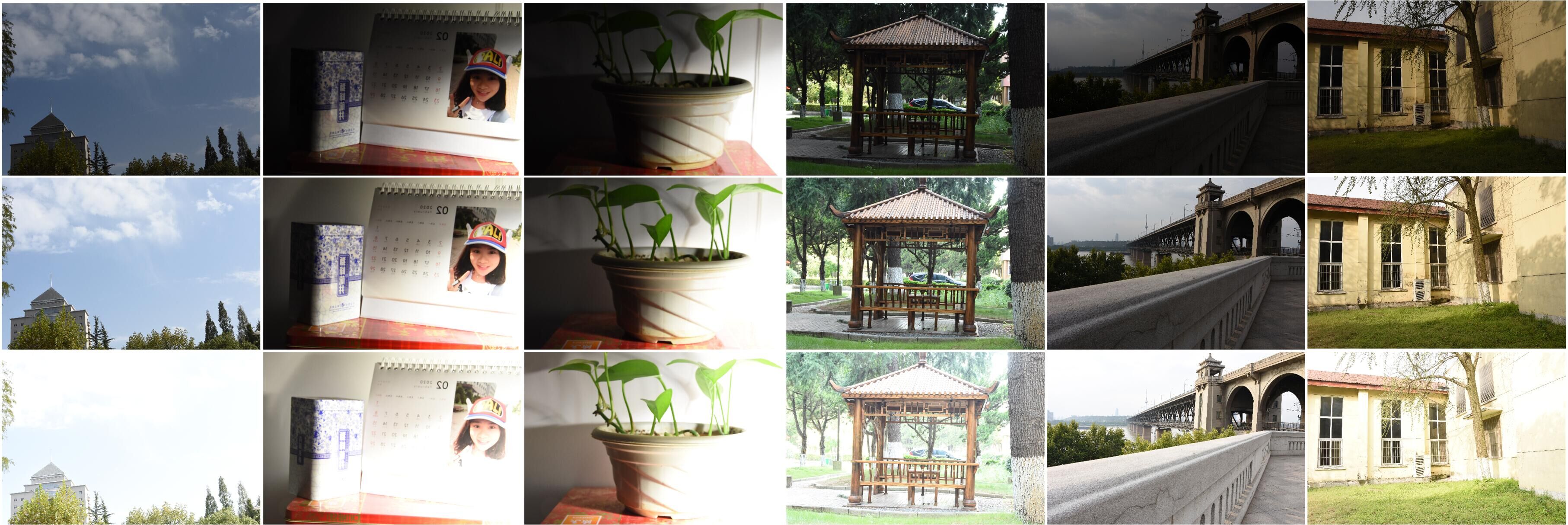}
	\caption{The first line are low exposure images. The second line are middle exposure images. The third line are high exposure images. The images are collected by changing exposure time, while other configurations of camera are fixed. The camera is fixed to mitigate the effects of jitter, and no moving objects can appear in the image, ensuring that the only variable is illumination. }
	\label{Fig.13}
\end{figure*}

Our datasets contains 370 multi-exposed image sequences. Each sequence has low/medium/high three images. Part of them are shown in Fig. \ref{Fig.13}. The interval of exposure ratio between them is 2 EV. Thus, the interval of two inputs is 4EV in the following experiments. The images are all captured by ourselves using Nikon 7200. To avoid other influence, only exposure times are changed while other configurations of the cameras are fixed. Also, Camera shaking, object movement are strictly controlled. Our datasets are diverse, including architecture, plants, daily necessities, etc., which meet the needs of DenoiseNet learning. Finally, we randomly split the images in the datasets into two subsets: 300 images for training and the rest for testing.

\subsection{Comparison of Two Different IMF Estimation Methods}

\begin{figure}[htb]
	\centering
	\includegraphics[width=0.450\textwidth]{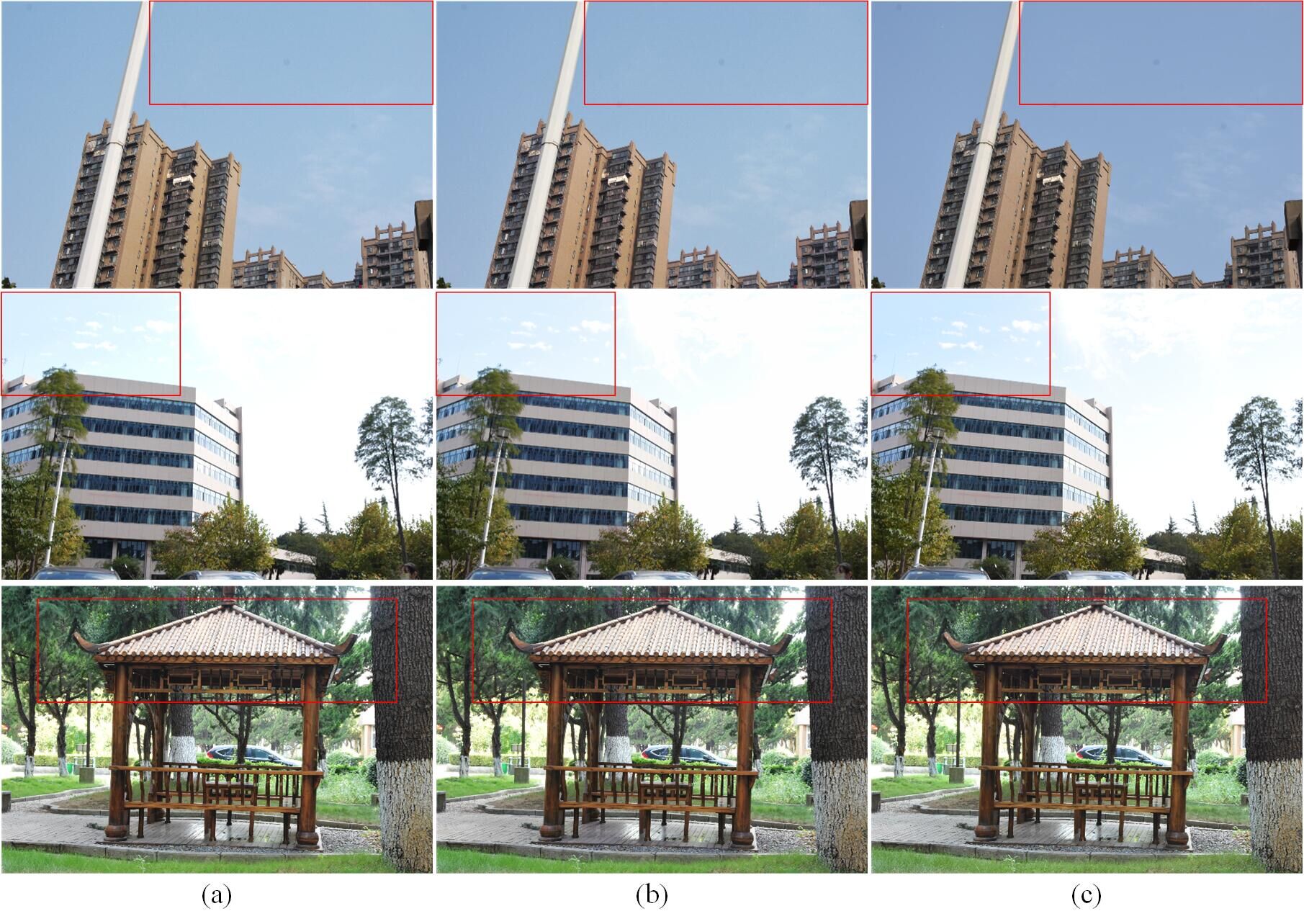}
	\caption{(a) The interpolated images via the method in \cite{1yang2018, 1zheng2020}; (b) The interpolated images via the proposed IMF estimation method; (c) The ground truth images. }
	\label{Fig8a}
\end{figure}

In this subsection, we compare the two different IMF estimation methods mentioned in section \ref{section:IMF} for estimating the continuous curve based on the  $F_c(\cdot)$ scatter plot. The $\Lambda_{i,3,c}(z)$'s obtained by the two methods are used as the inputs of Equation (\ref{newfusion}), and the interpolated image is compared with the ground truth. As shown in Table \ref{tabIMF}, by calculating the average SSIM and PSNR on 100 sets of test images, it can be objectively proven that the proposed IMF estimation method can interpolate  more accurate images than the method in \cite{1yang2018,1zheng2020}. At the same time, as shown in Fig. \ref{Fig8a}, the results of the proposed IMF estimation method look closer to the real image than the results by \cite{1yang2018, 1zheng2020}, which objectively proves the superior performance of the proposed method.

\begin{table}[htb]
	\begin{center}
		\centering
		\caption{SSIM and PSNR of tow different methods}
		\tabcolsep8pt\begin{tabular}{cccc}
			\hline		
			\multirow{1}*{ }   &   SSIM    &  PSNR\\
			\hline
			\multirow{1}*{IMF estimation method in \cite{1yang2018,1zheng2020} } & 0.9272 & 29.60\\
			\multirow{1}*{ Proposed IMF estimation method } & \textbf{0.9289} & \textbf{30.28}\\
			\hline
		\end{tabular}
		\label{tabIMF}
	\end{center}
\end{table}

\begin{figure}[!htb]	
	\begin{center}
		\includegraphics[width=3.0in]{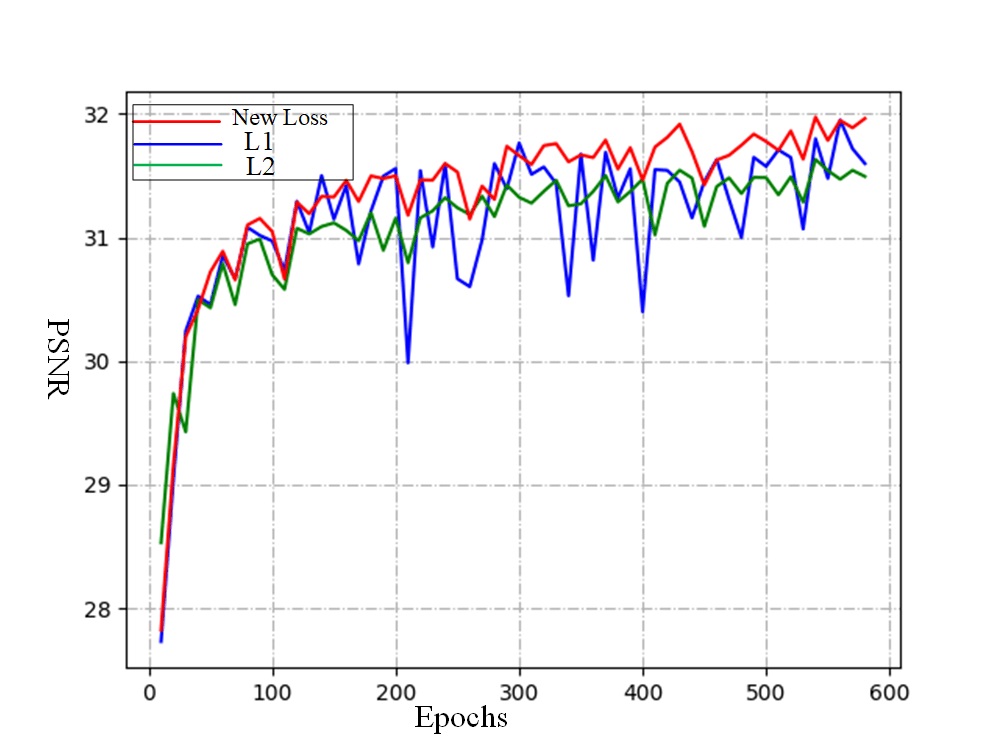}
	\end{center}
	\caption{Comparisons of the PSNR values among  the $L_1$ and $L_2$ loss functions (\ref{lossfunction1}) as well as the new loss function (\ref{lrlr}).}
	\label{Fig9}
\end{figure}

\subsection{Ablation Study on Loss Functions}
Since the main objective of this paper is to explore the hybrid learning framework rather than a more sophisticated neural network for deep learning, simple ablation study is conducted on the loss functions. It will show that the quality of the results will be improved with the proposed loss functions, even when the network architecture has not been changed.

\begin{figure*}[!htb]
	\centering
	\includegraphics[width=0.95\textwidth]{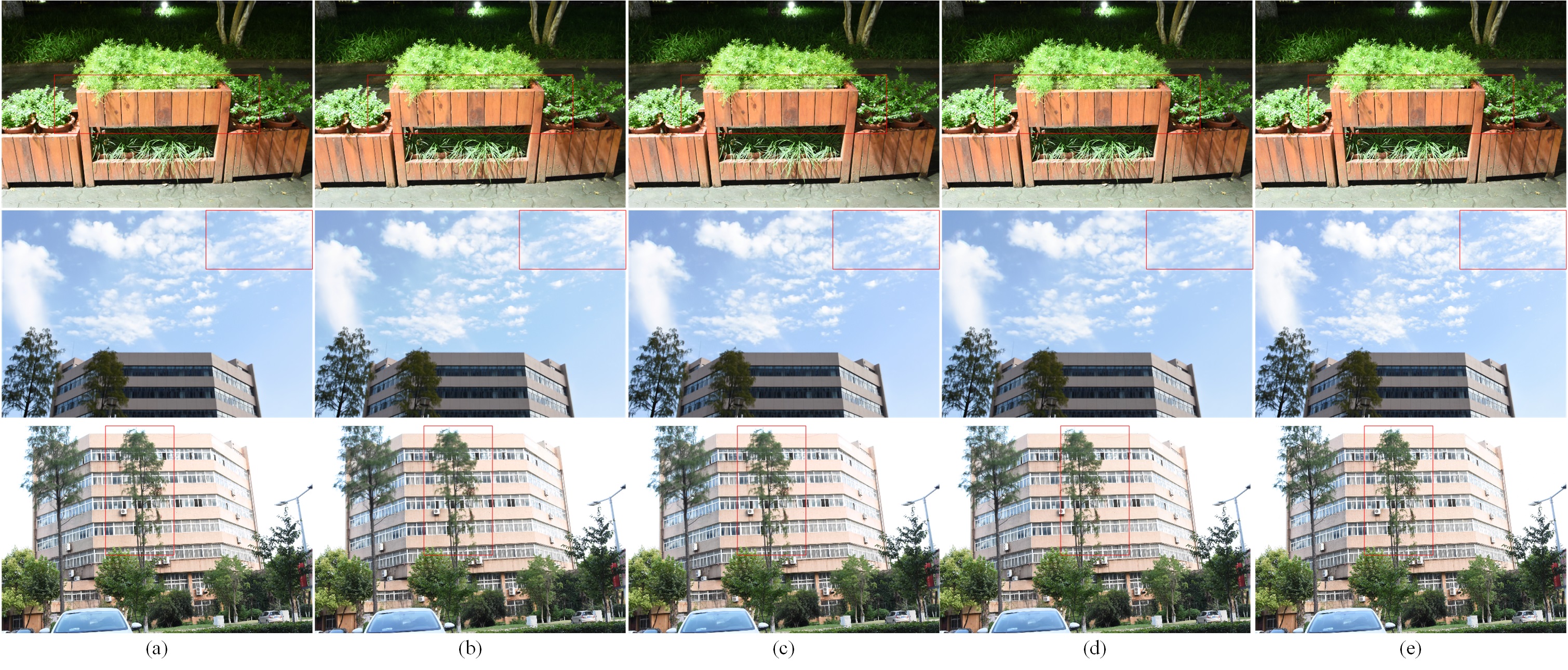}
	\caption{(a) are the results by using $L_r$. (b) are the results by using $L_r+L_c$. (c) are the results by using $L_r+L_f$. (d) are the results by using $L_r+L_c+L_f$. (e) are the ground truth images.  }
	\label{Fig10}
\end{figure*}

 The new loss function (\ref{lrlr}) is first compared with the $L_1$ and $L_2$ loss functions in (\ref{lossfunction1}). The values of PSNR for different epochs are shown in Fig. \ref{Fig9} and Table \ref{tab123}, the prosed hybrid $L_1/L_2$ loss function can obtain higher PSNR than both $L_2$ and $L_1$. Hence, the new loss function (\ref{lrlr}) is chosen as the loss function in the proposed method.

Although the restoration loss $L_r$ can implicitly measure the color difference, it cannot guarantee that $(f_0(x) + \tilde{f}(x))$ and $y$ have the same color direction. There may exist color distortion by using the restoration loss only, as shown in Fig. \ref{Fig10}. By adding the color loss $L_c$, the color distortion can be reduced. Both the $L_r$ and the $L_c$ are the pixel-wise loss functions which accurately capture the low frequencies but fail to encourage high frequency crispness. The resultant virtual image is high fidelity but not realistic. The image is usually overly-smooth
and thus has poor perceptual quality. Thus, feature-wise loss functions is applied to enhance the pixel-wise loss functions. As shown in Fig. \ref{Fig10}, the results of $L_r+L_f$ are much sharper than the results of $L_r$.  The SSIM and PSNR are also adopted to demonstrate the effectiveness of each component($L_r$, $L_c$ and $L_f$ ) and the result as shown in Table \ref{tab123}. Clearly, the images are also improved from the SSIM and PSNR points of view.

\subsection{Comparison of the Proposed Method with the Model-Driven Method}
In this subsection, the proposed framework is compared with the model-driven method in \cite{1yang2018}  to demonstrate the superiority of our algorithm from both the subjective and objective points of view.

As shown in Tables \ref{tabIMF} and \ref{tab123}, the average SSIM and PSNR values of 100 test images are much higher than those of the method in \cite{1yang2018}. This implies that the interpolated images by the proposed framework are much closer to the ground truth images than those by the method in \cite{1yang2018} from the objective point of view.
\begin{table}[htb]
	\begin{center}
		\centering
		\caption{SSIM and PSNR of three different choices}
		\tabcolsep8pt\begin{tabular}{cccc}
			\hline		
			\multirow{1}*{ }   &   SSIM    &  PSNR\\
			\hline
			\multirow{1}*{Proposed  $(L_1)$ } & 0.9374 & 32.23\\
			\multirow{1}*{Proposed  $(L_2)$ } & 0.9349 & 31.76\\
			\multirow{1}*{Proposed  $(new \ loss)$ } & 0.9399 & 32.31\\
			\multirow{1}*{Proposed  $(new \ loss+L_c)$ } & 0.9407 & 32.49\\
			\multirow{1}*{Proposed  $(new \ loss+L_f)$ } & 0.9406 & 32.42\\
			\multirow{1}*{Deep Learning $(new \ loss+L_c+L_f)$ } & 0.9444 & 32.69\\
			\multirow{1}*{Proposed $(new \ loss+L_c+L_f)$ } & \textbf{0.9455} & \textbf{33.32}\\
			\hline
		\end{tabular}
		\label{tab123}
	\end{center}
\end{table}

\begin{figure*}[!htb]
	\centering
	\includegraphics[width=0.95\textwidth]{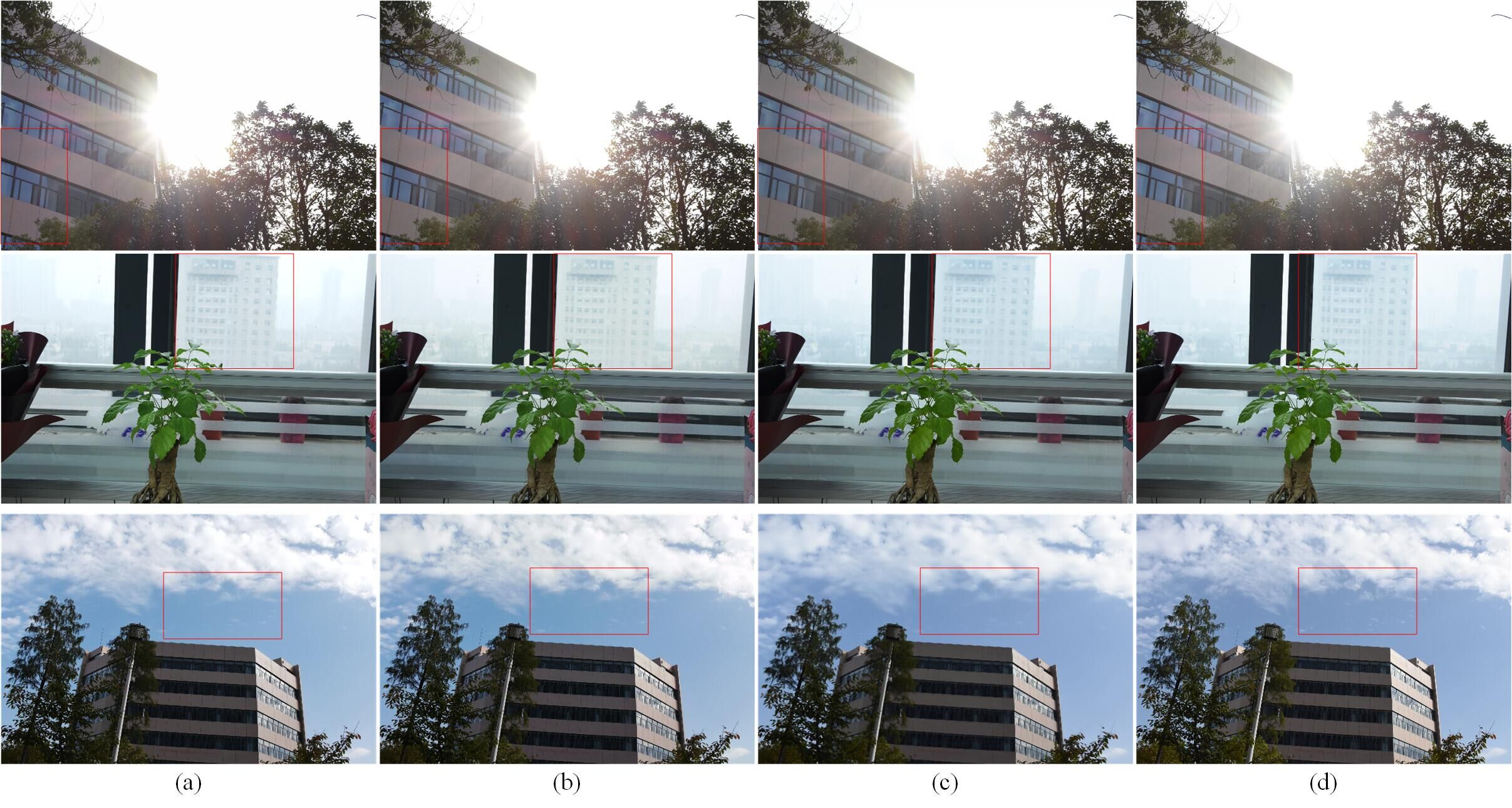}
	\caption{(a) The interpolated images $y_0$ via the method in \cite{1yang2018}. (b) The interpolated images by the proposed IMF. (d) The interpolated images by the proposed Method. (d) The ground truth images $y$; The proposed framework preserves more details than the method in \cite{1yang2018} without color distortion.}
\label{fig56}
\end{figure*}

The proposed algorithm is also compared with the method in \cite{1yang2018} from the visual quality point of view. As described above, the unmodeled information by the method in \cite{1yang2018} $(y-y_0)$ does exist. The proposed framework combines model-driven with data-driven methods to learn the residual image $(y-y_0)$. As shown in Fig. \ref{fig23}, the residual image $(y-y_0)$ by in the method in \cite{1yang2018} includes more visible information even though the pixel values are small but mostly non-zero. As shown in Fig. \ref{fig56}, the results by using the proposed method are much closer to the ground truth images than the images via the method in \cite{1yang2018} and the proposed IMF method.  These demonstrate that the proposed hybrid network can make up for the missing details in the image generated via model-driven exposure interpolation.

\subsection{Comparison of the Proposed Method with the Data-Driven Method}
\begin{figure}[!htb]	
	\begin{center}
		\includegraphics[width=3.5in]{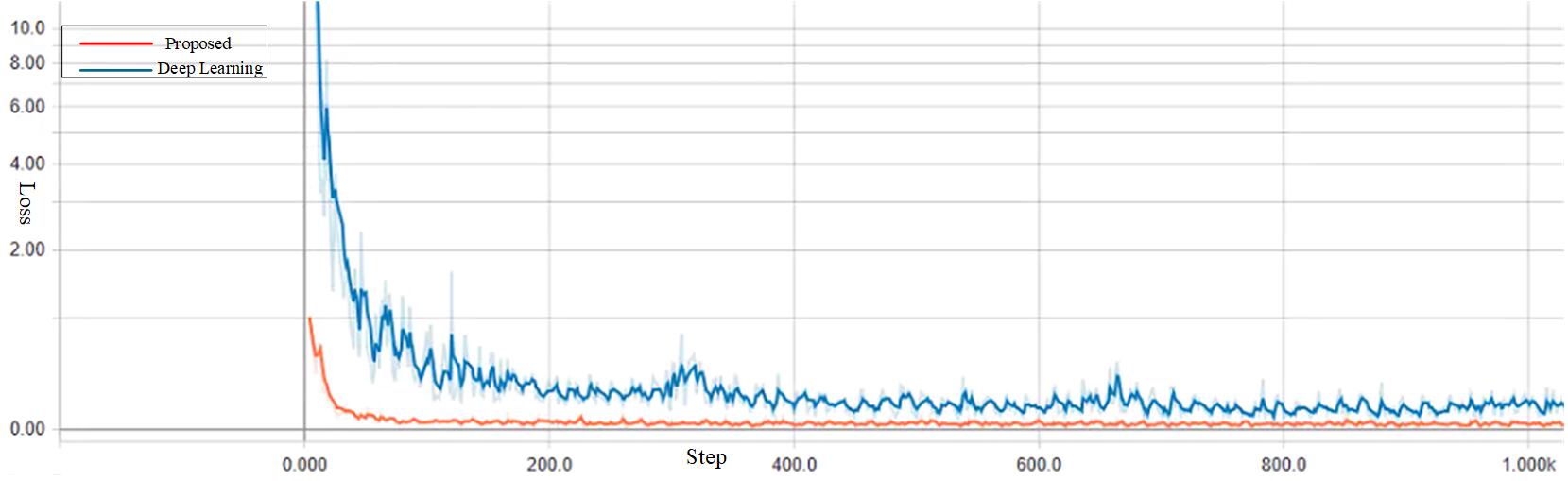}
	\end{center}
	\caption{Comparison of training, the red is the proposed hybrid framework, the blue is existing deep learning method.}
	\label{Fig7}
\end{figure}

 The structure of the deep learning method is shown in Fig. \ref{Fig6}, $x_1$ and $x_2$ are processed by two different convolutional neural network to generate the corresponding features, then the concatenated features are further processed by a convolutional neural network. The training convergence is shown in Fig. \ref{Fig7}. Obviously, the proposed solution converges faster and more stable than the alternative due to the desired outputs from our network are sparser and more convenient to be modeled through learning.

\begin{figure}[!htb]	
	\begin{center}
		\includegraphics[width=3.0in]{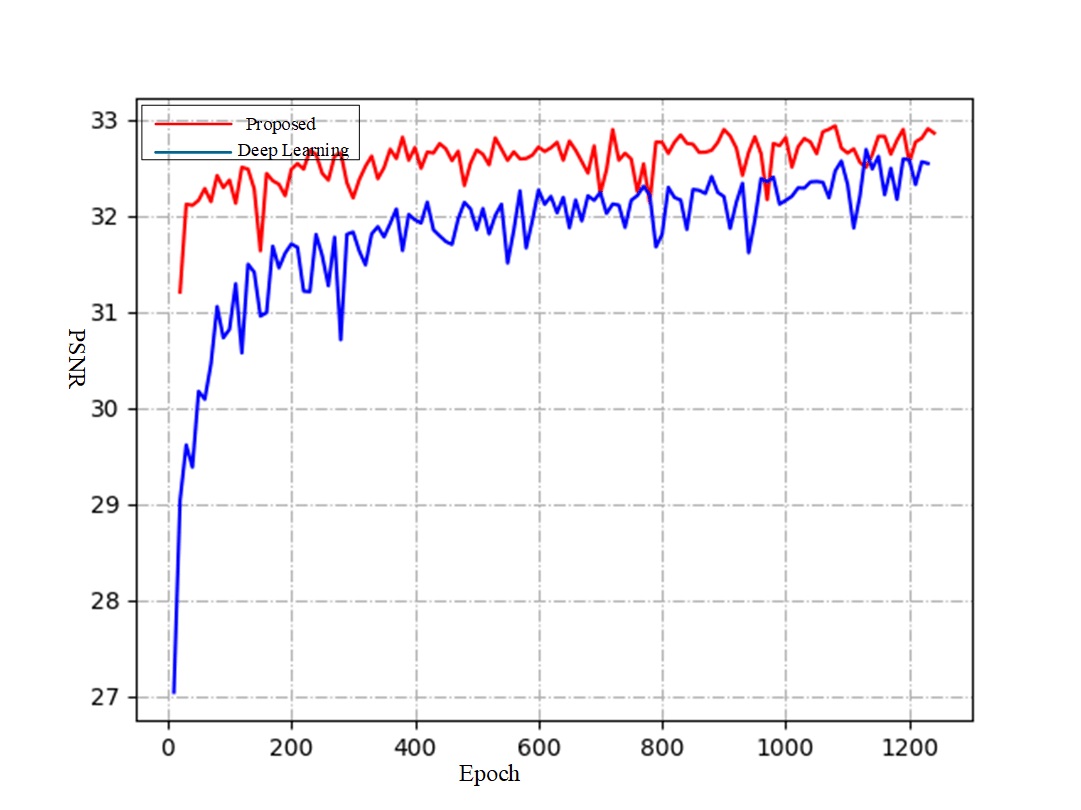}
	\end{center}
	\caption{Comparisons of PSNR between the proposed hybrid framework and a deep learning method, the red is our hybrid learning framework, the blue is deep learning method.}
	\label{Fig7.1}
\end{figure}

The quality of finally interpolated images by both methods with different iterations is shown in Fig. \ref{Fig7.1}. Clearly, our method converges much faster and more stable than the deep learning method from the PSNR point of view. As shown in Table \ref{tab123}, the proposed framework can obtain higher PSNR and SSIM results than deep learning method.

\subsection{Comparison with State-of-the-art MEF Algorithms}
As an application, the proposed method is adopted to improve multi-scale exposure fusion. Same as the algorithm in \cite{1yang2018}, our fused image is generated by fusing two different exposed images with one interpolated image by using the MEF algorithm in \cite{1mertens2007}. Here, five state-of-art MEF algorithms in \cite{1mertens2007,Ancuti2017,ZGLi2017,kou2017,1yang2018} are compared with our proposed method. It is worth noting that the input images of all algorithms are two true exposure images, whose exposure ratios are 16. All fused images are evaluated in terms of MEF-SSIM with the reference images as the three ground truth images with different exposure times.

\begin{figure*}[!htb]
	\centering
	\includegraphics[width=0.95\textwidth]{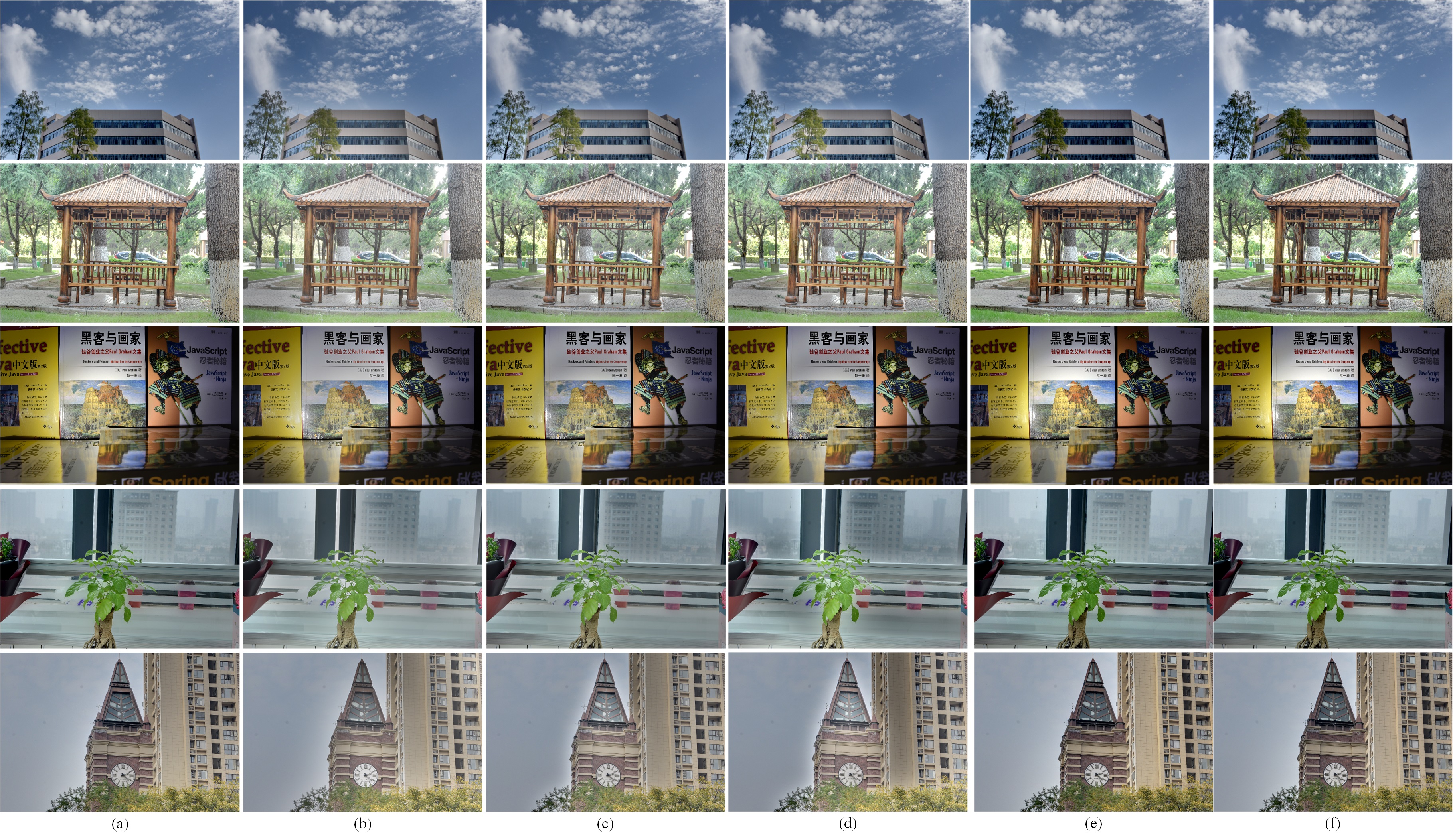}
	\caption{Results of six fusion algorithms. (a) fused images by using \cite{1mertens2007}; (b) fused images by using \cite{Ancuti2017}; (c) fused images by using \cite{ZGLi2017}; (d) fused images by using \cite{kou2017}; (e) fused images by using \cite{1yang2018}; (f) fused images by using our method. }
\label{Figfusion}
\end{figure*}

As shown in Table \ref{tabfusion},  the proposed algorithm outperforms all the six state-of-the-art MEF algorithms in terms of the MEF-SSIM. Part of the results are shown in Fig. \ref{Figfusion}. There are visible relative brightness reversal artifacts in the fused images by the algorithms in \cite{1mertens2007,ZGLi2017,Ancuti2017,kou2017}.  There are also visible halo artifacts in the last set of fused images by the algorithms in \cite{ZGLi2017,Ancuti2017,kou2017}. Although the results in \cite{1yang2018} can preserve the relative brightness order, some fine details are still missed. All these problems are overcome by the proposed method. Clearly, the exposure interpolation is indeed necessary for the fusion of two large-exposure-ratio images.

\begin{table}[t]
	\setlength{\abovecaptionskip}{0.cm}
	\setlength{\belowcaptionskip}{-0.cm}
	\caption{MEF-SSIM Of Six Different Algorithms}
	\centering
	\begin{tabular}{ c|ccccccccccccccccc }
		\hline
		\multicolumn{1}{c|}{} & \multicolumn{1}{c}{\cite{1mertens2007}}& \multicolumn{1}{c}{\cite{Ancuti2017}}& \multicolumn{1}{c}{\cite{ZGLi2017}}& \multicolumn{1}{c}{\cite{kou2017}}& \multicolumn{1}{c}{\cite{1yang2018}}& \multicolumn{1}{c}{Ours}\\
		\hline
		\multirow{1}*{Set1}  & 0.9633     & 0.9531     & 0.9606      & 0.9597    & 0.9653   &\textbf{0.9657}   \\
		\multirow{1}*{Set2}  & 0.9027     & 0.8834     & 0.8968      & 0.8975    & \textbf{0.9436}   &\textbf{0.9436} \\
		\multirow{1}*{Set3}  & 0.9454     & 0.9197     & 0.9416      & 0.9380    & 0.9641   &\textbf{0.9650} \\
		\multirow{1}*{Set4}  & 0.9200     & 0.9342     & 0.9356      & 0.9116    & 0.9571   &\textbf{0.9577} \\
		\multirow{1}*{Set5}  & 0.9670     & 0.9606     & 0.9654      & 0.9613    & 0.9734   &\textbf{0.9743} \\
		\multirow{1}*{Set6}  & 0.8920     & 0.8895     & 0.8966      & 0.8820    & 0.9310   &\textbf{0.9332} \\
		\multirow{1}*{Set7}  & 0.9594     & 0.9567     & 0.9601      & 0.9586    & 0.9741   &\textbf{0.9746} \\
		\multirow{1}*{Set8}  & 0.9715     & 0.9675     & 0.9721      & 0.9699    & 0.9780   &\textbf{0.9809} \\
	    \multirow{1}*{Set9}  & 0.9305     & 0.9356     & 0.9338      & 0.9279    & 0.9456   &\textbf{0.9459} \\
		\multirow{1}*{Set10} & 0.9525     & 0.9442     & 0.9546      & \textbf{0.9616}    & 0.9352   & 0.9443 \\
		\hline
		\multirow{1}*{Avg}  & 0.9404      & 0.9344     & 0.9417      & 0.9368    & 0.9568   &\textbf{0.9585}\\
		\hline	
	\end{tabular}
	\label{tabfusion}
\end{table}

\section{Conclusion Remarks and Discussion}
\label{conclusion}
A hybrid framework is proposed for exposure interpolation of  two large-exposure-ratio images by fusing a conventional method with a deep learning method. The deep learning method improves the quality of the intermediate image generated by the conventional method. The conventional method increases the convergence speed of the deep learning method and reduce the number of training samples required by the deep learning method. They {\it compensate} each other very well. All the interpolated image and the two large-exposure-ratio images are fused together via a multi-scale exposure fusion algorithm. Experimental results indicate that the exposure interpolation is indeed necessary for the two large-exposure-ratio images.

The proposed framework is scalable from the complexity point of view. It is attractive for ``capturing the moment" via mobile computational photography in the coming 5G era. The conventional method can be adopted to produce an image for previewing on the mobile device. The set of captured images will be simultaneously sent to the cloud and an image with a higher quality will be synthesized immediately. The synthesized image in the cloud will be sent back to the mobile device instantly due to the low latency of the 5G. If the photographer does not like the synthesized image, she/he can capture another set of images immediately.

\end{document}